# Towards a common performance and effectiveness terminology for digital proximity tracing applications

Justus Benzler, Robert Koch Institute, Berlin, Germany; BenzlerJ@rki.de

Dan Bogdanov, Cybernetica AS, Tallinn, Estonia; dan.bogdanov@cyber.ee

Göran Kirchner, Robert Koch Institute, Berlin, Germany; KirchnerG@rki.de

Wouter Lueks, School of Computer and Communication Sciences, EPFL, Switzerland; wouter.lueks@epfl.ch

Raquel Lucas: Medical School and Institute of Public Health (EPIUnit), Universidade do Porto, Porto, Portugal; rlucas@med.up.pt

Rui Oliveira, INESC TEC & University of Minho, Portugal; rui.oliveira@inesctec.pt

Bart Preneel, KU Leuven and imec, Leuven, Belgium; bart.preneel@esat.kuleuven.be

Marcel Salathé, Digital Epidemiology Lab, Global Health Institute, School of Life Sciences, School of Computer and Communication Sciences EPFL, Geneva, Switzerland; marcel.salathe@epfl.ch

Carmela Troncoso, School of Computer and Communication Sciences, EPFL, Switzerland; carmela.troncoso@epfl.ch (corresponding author)

Viktor von Wyl, Digital and Mobile Health Group, Epidemiology, Biostatistics and Prevention Institute, University of Zurich, Switzerland; viktor.vonwyl@uzh.ch (corresponding author)

*(authors in alphabetical order, all authors contributed equally)*

Correspondence to:

Prof. Viktor von Wyl, PhD
Epidemiology, Biostatistics & Prevention Institute
University of Zurich
Hirschengraben 84
CH-8001 Zurich
viktor.vonwyl@uzh.ch

Prof. Carmela Troncoso, PhD
EPFL IC IINFCOM SPRING
BC 258 (Bâtiment BC)
Station 14
CH-1015 Lausanne
carmela.troncoso@epfl.ch

# Abstract

Digital proximity tracing (DPT) for Sars-CoV-2 pandemic mitigation is a complex intervention with the primary goal to notify app users about possible risk exposures to infected persons. DPT not only relies on the technical functioning of the proximity tracing application itself and its backend server, but also on seamless integration of health system processes such as laboratory testing, communication of results (and their validation), generation of notification codes, manual contact tracing, and management of app-notified users.

Policymakers and DPT operators need to know whether their system works as expected in terms of speed or yield (performance) and whether DPT is making an effective contribution to pandemic mitigation (also in comparison to and beyond established mitigation measures, particularly manual contact tracing). Thereby, performance and effectiveness are not to be confused. Not only are there conceptual differences but also diverse data requirements. For example, comparative effectiveness measures may require information generated outside the DPT system, e.g. from manual contact tracing.

This article describes differences between performance and effectiveness measures and attempts to develop a terminology and classification system for DPT evaluation. We discuss key aspects for critical assessments of whether the integration of additional data measurements into DPT apps - beyond what is required to fulfill its primary notification role - may facilitate an understanding of performance and effectiveness of planned and deployed DPT apps. Therefore, the terminology and a classification matrix may offer some guidance to DPT system operators regarding which measurements to prioritize.

DPT developers and operators may also make conscious decisions to integrate measures for epidemic monitoring but should be aware that this introduces a secondary purpose to DPT that is not part of the original DPT design. Ultimately, the integration of further information (e.g. regarding exact exposure time or notification chains and their delays) into DPT involves a trade-off between data granularity and linkage on the one hand, and privacy on the other. More data may lead to better epidemiological information but also increase the privacy risks associated with the system, and thus also decrease public DPT acceptance. Decision-makers should be aware of the trade-off and take it into account when planning and developing DPT notification and monitoring systems or intending to assess the added value of DPT relative to existing contact tracing systems.

# Introduction

Digital proximity tracing (DPT) is a novel health technology, designed to complement manual contact tracing (MCT) by using apps in national efforts to mitigate the Sars-CoV-2 pandemic.[1] *The primary purpose of DPT is to provide an instrument for fast, anonymous notification of other app users with potential exposure risks to an infected app user. [2]*
The current Sars-CoV-2 crisis is the first global public health crisis that sees massive, nationwide roll-outs of DPT apps.[3] This is noteworthy because DPT has never undergone large-scale, real-world testing in its target population, as would normally be required for health technologies.[4,5] This fact is largely owed to the urgency of the Sars-CoV-2 crisis, where many countries gave precedence to fast release schedules over extensive testing. All the more, governments face pressure to justify the rapid DPT deployment and to demonstrate its impact on pandemic mitigation. Specifically, evaluations are needed to demonstrate whether single parts and the whole system of DPT perform well from a technical perspective, but also whether DPT helps to contain transmission chains.[6]

Ideally, such evaluations of DPT should follow a standardized protocol to allow comparability across countries and settings, but also to facilitate learning from other countries' experiences. However, in direct exchanges and discussions with national health authorities or developers from other countries, this group of authors noticed a substantial confusion among health authorities, politicians, and even DPT experts about the aims of DPT, the terminology, and goals for system evaluations. The metaphor of a "Babylonian confusion of tongues" is not too far to describe the current situation. This problem has been recognized, and international health authorities and different groups of academics have attempted to bring some structure into discussions about DPT development, deployment, and evaluation.[7,8]

The present viewpoint attempts to provide further clarifications on key aspects of DPT evaluations by bringing together DPT developers and public health experts from different countries to present a unified proposal for a terminology and classification of measures to evaluate DPT. The viewpoint is structured as follows. Section 1 describes the basic principles of DPT. Section 2 argues that DPT is a complex intervention, relying on the fast completion of clearly defined actions in the notification cascade by different health systems actors. Section 3 breaks the DPT notification cascade into its separate parts and describes how some basic questions and checks may easily lend themselves to DPT evaluation. Section 4 introduces basic concepts and terminologies to describe and assess DPT systems from different viewpoints, namely system performance assessments and public health effectiveness evaluations. Section 5 outlines a classification matrix to distinguish different types of indicator measures. In section 6, the viewpoint closes with some basic considerations for developing and implementing indicators for DPT evaluation.

# 1 Principles of digital proximity tracing in support of manual contact tracing

The principles of DPT have been described extensively elsewhere.[1] In brief, DPT enables participants to trace proximity contacts (exposures) that could pose relevant infection risks. If one of the proximity contact persons tests positive for Sars-CoV-2, the app will warn other users who were in close proximity within the infected person's time window of infectivity, thereby increasing the coverage and/or speed of the contact tracing process relative to MCT. The potential advantages of DPT compared with MCT are threefold. DPT can (1) lead to faster exposure notifications than MCT, DPT can (2) reach persons who are not personally known to an index case, and DPT (3) is easily scalable and should still work when MCT reaches its capacity limits.

Most countries that have released DPT apps have opted for a privacy-preserving, decentralized architecture according to the DP-3T blueprint.[9,10] That is, proximity contacts are not sent to a central server, but stored and evaluated locally on smartphones. The only data that is sent to a central server are pseudonymous, random identifiers of persons with a confirmed SARS-CoV-2 infection. These "infectious" identifiers are downloaded by all other users and compared to the locally stored identifiers to find out which of the proximity encounters were with SARS-CoV-2 positive people. If the temporal aggregation of the matched encounters exceeds a minimal duration at a relevant proximity (depending on the estimated infectiousness of the positively tested person), users are notified and recommendations on further steps to take are provided.

The dominant choice of a privacy-preserving architecture by many countries highlights the emphasis of the primary DPT core function ("warning people early in an anonymous manner") over purposes such as disease monitoring. DPT is, first and foremost, a notification tool aimed at breaking transmission chains, and its primary function does not necessitate the collection of personal data of index cases and their contacts. Nevertheless, debates in several countries suggest that DPT is sometimes also viewed as an opportunity to collect data for epidemiological monitoring, for example, to obtain additional information on time and setting of risk exposure events. Such *secondary functions* of DPT are beyond the scope of this article and are only discussed briefly where relevant for the broader context.

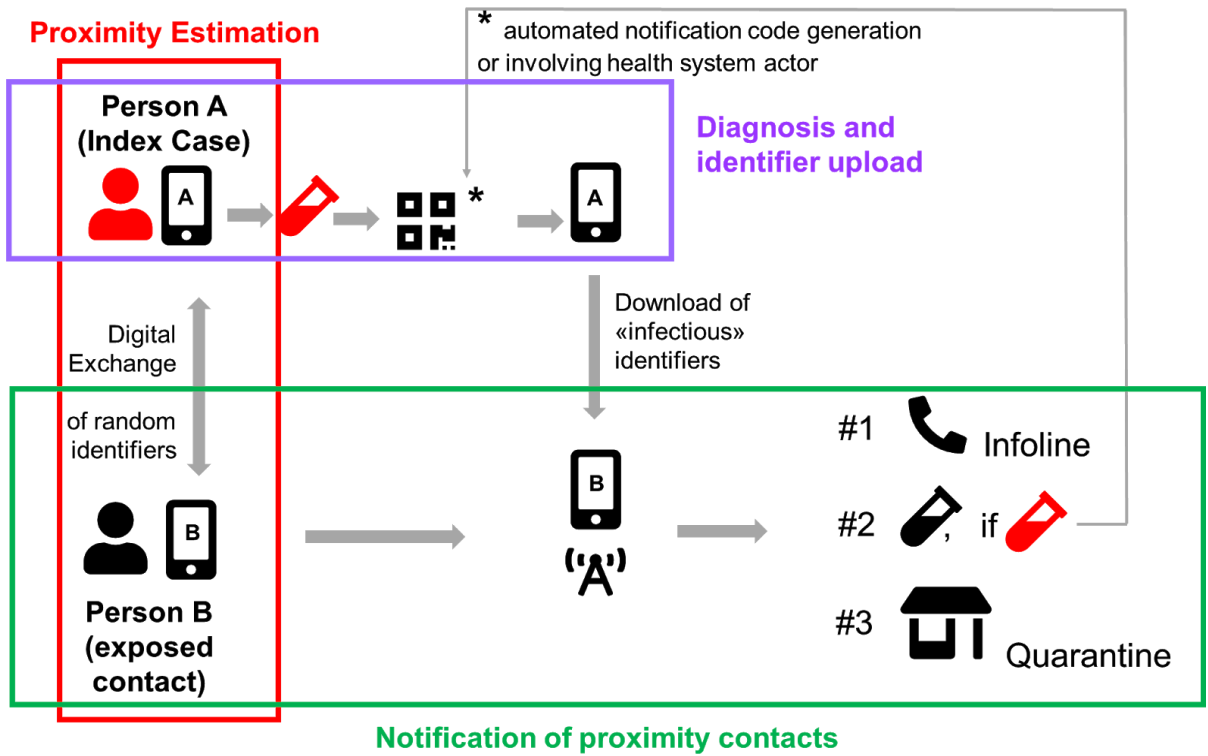

*Figure 1: Example of required steps in the notification cascade of digital proximity tracing in decentralized systems. An infected person A receives a positive test result for Sars-CoV-2 and (possibly automatically) an upload authorization. After consent by the user, the app uploads the random identifiers to a server. Person B's device regularly downloads "infectious" identifiers. If B was in close proximity to A (or other infected users) for a prolonged time, then B receives the app notification. Upon receiving this notification, B has several options, for example - in some countries - calling an information line (#1), getting tested for Sars-CoV-2 (#2), or entering quarantine voluntarily (#3). The colored boxes refer to the three main tasks involved in the DPT notification cascade, described in section 2.*

# 2 DPT is a complex intervention

The preventive effect of DPT results from a timely warning of exposed persons so that they can enter quarantine and initiate further preventive measures. In this context, timely means a faster contact notification than is usual in MCT. In addition, DPT can also reach persons who would normally be missed by MCT (e.g. because they were chance encounters of the index case). As illustrated by Figure 1, the app notification process reflects an information flow in multiple steps to eventually produce specific actions leading to the prevention of further transmission (indicated by the #-sign in Figure 1). The effect of DPT depends on the interplay between health system actors (e.g. testing laboratories) and *app users*. It is not the app per se but the fast completion of the full notification cascade and subsequent actions that lead to the desired results.[11] Of note, the distinction between app users and other actors is warranted because (voluntary) user actions are strongly influenced by behavioral aspects and incentives [12,13], whereas actions required by other health system actors may depend more on automatization, technical interfaces, resources, or capacity [14].

We identify three high-level tasks (illustrated by colored boxes in Figure 1) that combined cover the entire notification cascade:

- **Proximity estimation.** This aspect pertains to the exchange of ephemeral (regularly changing) identifiers between users' devices, and the detection of significant proximity contacts (e.g., less than 1.5 meters for more than 15 minutes).
- **Diagnosis and identifiers upload.** This aspect pertains to the upload of identifiers by index cases.
- **Notification of proximity contacts.** This aspect pertains to the notification of proximity contacts by their mobile device, and the subsequent actions taken by users as a response to this notification.

The dependency of the DPT intervention on its embedding in the overall pandemic mitigation response and the involvement of multiple actors fulfills the definition for complex interventions, as used in other fields of healthcare research.[15,16] In DPT, involved health system actors are setting-specific but may include testing laboratories or health authorities including MCT units or operators of infolines (Figure 1). Figure 2 provides an even more detailed view of ten individual steps in the DPT notification cascade. The red person illustrates the infected app user who gets tested, receives a positive test result, and triggers the app notification. The green person depicts a proximity contact who receives an app notification. Of note, in most DPT implementations, several of the steps in Figure 2 involve free user choices whether or not to complete a specific task (e.g. step 6, authorization of key upload) without fears of retribution.

This insight that DPT is a complex intervention involving voluntary actions is relevant from a practical standpoint because it shifts the focus of discussion from single aspects (e.g. technical accuracy of Bluetooth measurements) to a broader systems perspective.[17] But in addition to measuring the final intended outcome of a complex intervention, the monitoring of individual components of the complex intervention is nonetheless important. Because complex interventions, and DPT in particular, depend on a seamless, fast cascade of events (as shown in Figure 1), measurements characterizing speed and efficiency of specific system components and actors are useful to identify bottlenecks in the notification chain, as well as to act as leverage points for improving system behavior.

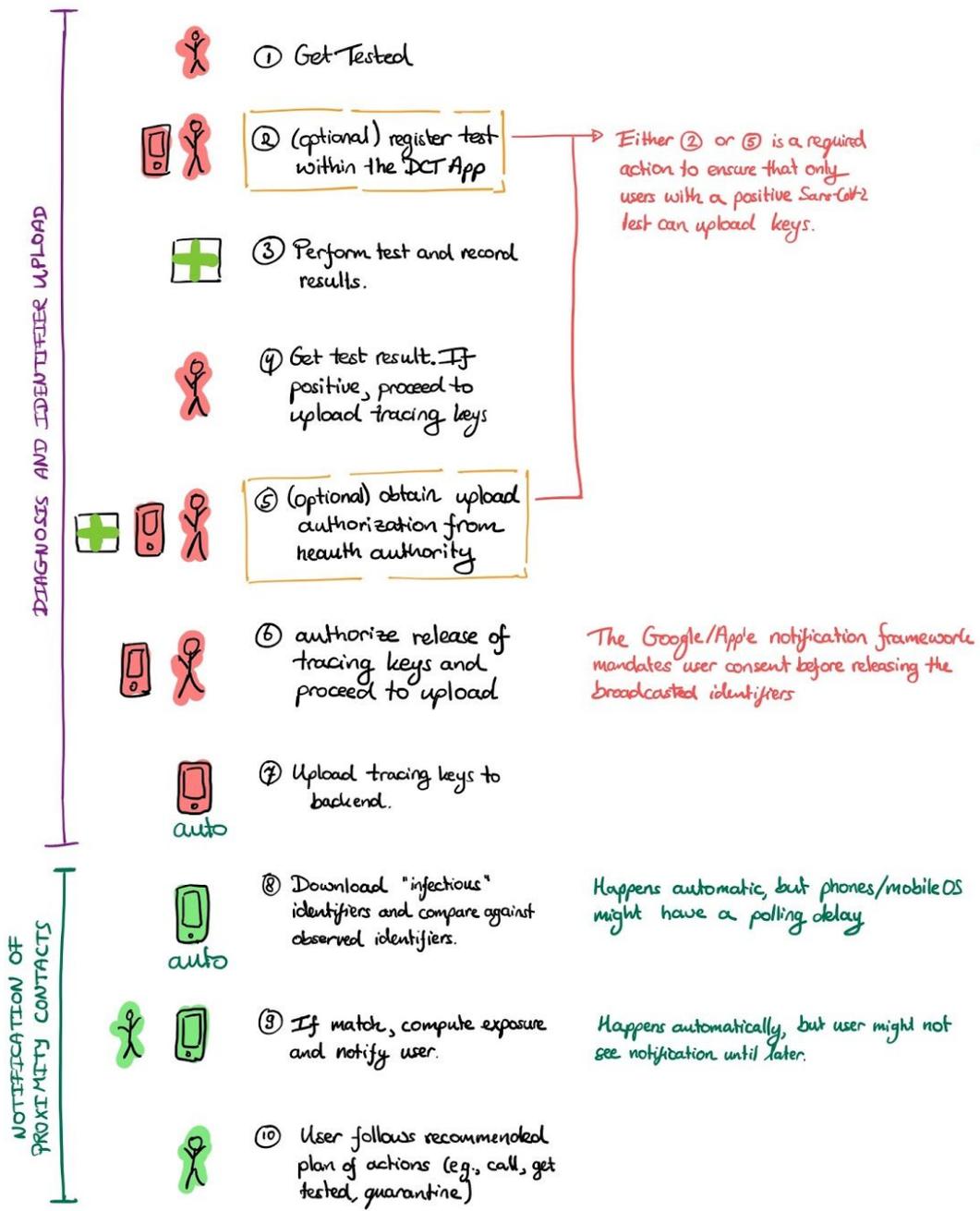

*Figure 2: Specific actions required in the notification cascade.*

# 3 A closer look at DPT steps and their influence on intervention outcomes

To illustrate how system components can influence the outcome of the intervention, let us look at the three high-level tasks indicated in Figure 1 above (*proximity estimation, diagnosis and identifiers upload, proximity contact notification*). The precise details of each of the tasks depend on national or regional choices regarding system design and configuration. However, all systems adhere to a similar structure based on the three steps (proximity estimation, diagnosis and identifier upload, and proximity contact notification). In the following, we will describe the processes in greater detail, describe country-specific variations, and identify questions that help to assess the system performance.

## 3.1 Proximity estimation

All systems we consider in this paper rely on the Google/Apple Exposure Notification (GAEN) framework.[18] Therefore, the accuracy of proximity estimation depends on the functioning of the GAEN framework and how well the chosen parameters reflect the desired measure of proximity (with slight regional variation).

Determining the accuracy of this component requires accurate ground-truth information about the exact distance and duration of a proximity contact. Collecting such information is nearly impossible in non-experimental settings without infringement of privacy (e.g., because it would require the use of video cameras to establish ground-truth). Therefore, research groups [19–22] have used laboratory and simulated settings to replicate these scenarios with carefully constructed/measured ground truth information. These measurements can then be used to answer questions such as:

- What parameter choices best reflect the desired distance/duration threshold?
- What distribution of false positives / false negatives does this choice induce for particular users' behaviors and under different environments?
- Does the choice of device model/manufacturer/platform influence the realized threshold?

To address these specific technical questions and to inform setting-specific implementation decisions, the GAEN framework documentation is a suitable starting place. [9]

## 3.2 Diagnosis and identifier upload

To enable proximity contacts to receive notifications, Sars-CoV-2-positive users must upload the random identifiers they broadcasted during the contagious window. Most countries use a system of **upload authorization** to allow uploads by confirmed index cases only to prevent false warnings or manipulation. Automatic uploads of identifiers are not possible, as of today.

The following critical steps are critical (Figure 2):

1. Get tested and receive a (positive) test result
2. Obtain upload authorization by a health authority (e.g., automatically when users registered the test in the app; or via an upload authorization code obtained through interaction with the health authorities after receiving a positive test)
3. Upload identifiers by consenting to the release of identifiers and uploading them in the app

The exact process of obtaining upload authorizations differs per country. Some countries, e.g., Belgium [23], and Germany [24], let users register Sars-CoV-2 tests in the app (who then also receive their test result through the app). Users with positive tests then automatically receive an upload authorization. In addition to or instead of such an automatic flow based on registered tests, other countries provide index cases with an upload authorization code that they enter into the app (Switzerland and Portugal use this as the only flow, German and Belgium use it as an alternative) or let users obtain authorizations via an eHealth system (e.g., in Estonia where the eHealth system also manages Sars-CoV-2 tests).

Moreover, the GAEN framework requires users to explicitly consent to release the random identifiers their phones have broadcast. Therefore, any app will request user action and explicit consent. However, the timing of consent provision varies across countries and can occur, for example, during laboratory test registration (e.g. in Germany) or upon receipt of a positive test result.

Based on the three critical processes described above, one could ask the following questions related to the operation of the system:
- How long between users getting tested and receiving a result?
- What fraction of users register tests in the app, if this function is available?
- What is the time from receiving the result to uploading identifiers?
- How many users that have permissions to upload do not finish the process? At what point do they abandon the process?

While data collections to address these questions are highly setting-specific, publicly accessible examples of such monitoring systems already exist, for example, in Switzerland [11,25], Germany [26] , or in the Netherlands [27].

## 3.3 Notification of proximity contacts

Apps regularly retrieve uploaded positive identifiers from the system's backend server and then compare them (using the GAEN framework) against stored identifiers. If the framework detects a sufficiently high exposure, the app sends a notification to the end-user. Depending on the country, this notification instructs the user to self-quarantine, take a test, or contact the local health authority or a hotline. As with manual contact tracing, users may fail to follow these instructions.

To protect privacy (and to facilitate efficiency), uploaded identifiers are not immediately downloadable by other app end-users. Instead, the backend regularly releases a new batch of identifiers. For example, every two hours.

The following factors influence the notification timeline.
1. Time to publish. The time between receiving "positive" identifiers at the system's backend, and the publication time where these identifiers can be retrieved by other phones.
2. Polling frequency. The frequency with which user's devices retrieve new "positive" identifiers and compute matches. This time is influenced by the mobile OS's scheduler as well as the internet connectivity of the phone.
3. Time for the user to notice the notification and to subsequently act upon it.

The first and second factors are configurable parameters, constrained by the capabilities enabled by the GAEN framework.[9] The effect of the system's scheduler can be tested in a laboratory setting.

The following questions can be asked concerning this task (some may not be possible to answer in privacy-preserving architectures):
- How many people receive a notification? How many of them later get positive?
- How many people follow through after receiving a notification?
- What is the time between positive upload and notification?
- How long between a notification is received and the user acts upon it?

The monitoring websites mentioned in section 3.2 also provide good example metrics for the proximity contact notification step. [25–27]

# 4 Basic concepts for DPT evaluations

As illustrated above (Figures 1 & 2), DPT relies on a fast and seamless flow of information along the notification cascade. Blockades or inefficiencies in single steps can lead to bottlenecks and prevent the information flow, thus inhibiting the primary goal of DPT to warn other app users about potential risk exposures. The information flow of each DPT step can be described by at least three attributes: *speed, yield, and capacity*. *Speed* describes how fast an action is completed and can be measured in terms of time. *Yield* refers to a completion fraction, that is, how many times per 100 a task was executed as needed. Yield sometimes also has a time connotation, that is the fraction of task completion for a given time frame (also referred to as *throughput*). *Capacity* relates to the task volume an actor can handle in a given time (which, in turn, may also influence speed and yield). For example, testing laboratories or manual contact tracers can only process a certain number of samples or cases, given the available resources such as machinery or personnel. Therefore, volumes can be described as percentages below or above the capacity limit.

The different DPT system components, as well as the basic attributes (*speed, yield, capacity*) and questions about DPT functioning, lend themselves to the development of *Key Performance Indicators* (KPI). For a better interpretation of these KPI, a contextualization with the dynamic of the Sars-CoV-2 pandemic is often useful (e.g. to compare the number of key uploads with the incidence of new infections).

Overall, such KPIs provide valuable information on the procedural performance of the overall system, as well as possible bottlenecks in the notification cascade. If any of the components of the tasks in Figure 1 (e.g. providing upload authorization) malfunctions, the delay will ripple to the whole notification cascade and undermine DPT's ability to perform its primary purpose of notifying exposed contacts. Therefore, KPIs can also be viewed as measures of technical and procedural preconditions for DPT to fulfill its primary purpose.[11]
However, KPIs have the caveat that they often summarily reflect sequences of different actions (e.g. app usage, positive test, identifier uploads, download of identifiers, and proximity estimation). The complexity of this composition hinders the interpretation of these metrics as a consequence of one factor.

Furthermore, KPIs are, in a strict sense, **not revealing** concerning how well DPT achieves its primary purpose of reducing viral transmissions, respectively its "*effectiveness*", which is defined as the "ability to produce a desired result".[28] In epidemiological studies, the concept of effectiveness often stands for the real-world effect of an intervention against a comparator (comparative effectiveness) and is expressed as an exposure-outcome relationship.

It is important to note that, in the majority of countries, DPT is intended and designed to complement MCT. Therefore, in most settings, a "fair" DPT effectiveness evaluation would involve a comparison between the use of DPT apps versus non-use in addition to manual contact tracing. In those settings where DPT is implemented alongside MCT, effectiveness investigations should ideally center around the three postulated main advantages of DPT over MCT: DPT should lead to faster exposure notifications than MCT, DPT can reach persons who are not personally known to an index case, and DPT should still work when MCT reaches its resource limits.[6] Specific effectiveness outcomes could focus, for example, on the time from first exposure notification (either by MCT and/or DPT) to entering quarantine, as well as comparisons of the average number of persons who later test PCR-positive between groups who were notified by MCT or DPT. In both examples, the most obvious comparator is classic MCT. In settings where no MCT exists (or where it is no longer operable) and DPT is introduced in a staggered process, regional comparisons of Sars-CoV-2 incidence could be performed between geographic units that have introduced DPT at different times.

Sometimes DPT and DPT-related measurements are also discussed in the context of *epidemiological monitoring* (surveillance). Such discussions are tied to the hope to gather relevant data and gain insights about transmission dynamics. It is important to note that epidemiological monitoring is not part of the key functionality of DPT and necessitates an entirely different set of measurements and KPIs that go beyond the data requirements for privacy-preserving proximity tracing; it is not included in the following discussions.

# 5 Proposal for a classification of different DPT evaluation measures

## 5.1 High-level distinction between key performance indicators and public health effectiveness metrics

Figure 3 proposes a *KPI classification matrix* of different measure types and perspectives ("aspects") relevant for DPT assessments. We acknowledge that the distinction between the different proposed types may not always be clean-cut in practice. In fact, one may need several of these metrics to assess how well the system performs each of the three tasks DPT systems must realize to fulfill their objective (see Figure 1).

The horizontal dimensions of the classification matrix show different steps from a basic Input-Processing-Output (IPO) model perspective.[29] Each step in the DPT notification cascade requires resources (e.g., technical infrastructure, money) and inputs (e.g. information), which are processed to create outputs (e.g. notifications). The different IPO steps can be examined from different viewpoints shown in the vertical classification matrix dimension, namely from a technical (app-) perspective, as well as from the viewpoint of different actors, which include app end-users, but also laboratories or public health services (Figure 1).

Therefore, each matrix cell reflects a combination of IPO-step and viewpoint that can be useful to describe and evaluate the performance of specific steps or elements of DPT systems (KPIs). By taking a specific step in the notification cascade described in section 3, the performance can be evaluated from different angles using the guiding questions such as: What are the resources needed to complete this step? How well does the information flow along the notification cascade work? Or how much desired output is generated by this step? Such KPIs can be formalized as raw numbers, proportions, or ratios to describe *speed, yield, and capacity* attributes (as described in section 4).

Separate and located below the classification matrix in Figure 3 are the public health *effectiveness* measures. They are distinct from KPIs and aim to address a different question: does the DPT system achieve its intended primary aim of notifying exposed app users swiftly so they can take preventive measures? Measures of DPT effectiveness can relate, for example, to the prevention of further transmission or comparative cost-effectiveness when compared to MCT.[6] The effectiveness measures are different from process KPIs, and yet not independent. The completion of the notification process is a precondition for achieving the DPT public health goal. In other words, many public health metrics are an integral of different processes in the app notification cascade, as they are a direct consequence of how effective the notification cascade tasks are executed.

|  | Resource-oriented metrics (What are resources and costs of specific DPT steps?) | Process-oriented metrics (Do processes built around specific DPT steps work as expected?) | Output-oriented metrics (How frequent are desired outputs of specific DPT steps?) |
|---|---|---|---|
| **App-oriented, technical perspective** (Do technical aspects of specific DPT steps work as expected?) | 1. Costs to develop and maintain DPT notification infrastructure (per user). | 3. Frequency of infectious identifier downloads. | 5. Turnaround time of test results for notified app users. |
| **Actor-centric perspective** (Are *app end users* behaving in specific DPT steps as expected? Are *key actors* engaging in specific DPT-steps as expected?) | 2. Costs to provide app notified contacts a free Sars-CoV-2 test. | 4. Time for exposed end user to notice warning. | 6. Fraction of notified end users entering quarantine. |

**Public health-centred effectiveness metrics**
(Is DPT achieving its primary public health goal against a comparator?
*real-world comparative effectiveness*)

*Figure 3: Proposal for classification matrix of different indicator types to monitor DPT systems by steps of the Input-Process-Output model (resource-, process-, and output metrics) as well as by aspect (app-oriented, technical viewpoint vs. actor-centric perspectives). The examples in the matrix pertain to the notification step of proximity contacts (section 3.3).*

## 5.2 A worked example of the KPI classification matrix and public health measures: the proximity notification step

To further illustrate the use and usability of the KPI classification matrix, we will – cell by cell – describe KPI examples related to the proximity notification step (which could also be applied to classic MCT). The proximity notification step (section 3.3) is a crucial element in the notification cascade with a direct relation to the primary DPT goal: to warn proximity contacts as early as possible about potential transmission risk exposures.

The first matrix column represents <u>resource-oriented metrics</u>, which define resource needs for technical and non-technical implementation and include, for example, costs for PCR-tests (which are free for persons with a DPT notification in some countries), costs for quarantining of DPT-notified persons, or any other expenditures.

In the vertical dimension, the first matrix row reflects the <u>app-oriented, technical perspective</u>. During DPT development and operation, IT system design requires choices for parametrization of measurements and backend systems, which are resource-dependent and may impact speed, yield, or capacity of DPT processes. Therefore, concrete KPI examples for resource-oriented metrics from the technical perspective (*cell 1*) include costs for development and maintenance of the app itself, as well as for the technical infrastructure and the backend. Such expenditures are often scaled by the number of active users or the number of quarantine orders (or other indicators for prevented transmissions).

The second matrix row reflects the actor-centric perspective on the notification chain. Involved actors include laboratories that perform PCR-tests and communicate results to app end-users and health authorities, or - in settings with manual upload code generation – the health authorities and other authorized persons who release upload keys or take calls from notified users. Furthermore, among all involved actors, the app end-users play a very central role. In some settings with lower degrees of notification cascade automatization (e.g. in Switzerland), much responsibility falls on end-users who need to decide whether to use the DPT app, but also to actively trigger (or at least consent to) the upload of identifiers in case of testing positive for Sars-CoV-2. Example KPIs that combine resource and actor perspectives (*cell 2*) include expenditures for user-linked actions, such as the costs for providing app-notified contacts a free Sars-CoV-2 test.

Process metrics are located in the second matrix column. Those KPI describe interactions of the app and its users with other parts and actors of the health system. For the app to work as intended, several processes need to occur seamlessly so that all tasks can be carried out successfully and timely: from testing to prompt results communication, notification code generation, and identifier upload, notification of exposed contacts, and these contacts taking action (e.g., calling the hotline or a doctor and receiving advice). Process metrics can be used to monitor how well the different conditions for app-functioning are met, respectively, whether the different system parts work as expected.

Examples of process-centric metrics that integrate the technical perspective (*cell 3*) include, for example, precision and recall of Bluetooth and exposure time measurements, which are usually assessed in experimental settings. Specific design choice evaluations may involve measurements of how well the GAEN/Bluetooth approximation reflects actual physical distance and time exposure, as well as backend configurations regarding the frequency of infectious key uploads or downloads of lists of infectious identifiers (which only happens a certain number of times per day). *Cell 4* represents process-oriented metrics from an actor perspective. Several steps in the notification cascade require human involvement, sometimes on a voluntary basis. Therefore, such actions may be strongly influenced by behavioral aspects, digital and health literacy, but also by incentives. KPI examples include the fraction of app users who consent to or actively initiate the identifier uploads in settings where this is voluntary and requires user actions. KPIs to describe such steps can often be based on yield (fraction of completed tasks) or speed (time to completion) attributes.

Finally, the third matrix column reflects output-oriented metrics. These KPI refer to desired outputs of DPT, which could be numbers or yields of DPT-notified users who undertake a recommended action (e.g. entering quarantine or getting tested for Sars-CoV-2). These metrics are related to public health goals but differ in that they focus on an intra-system perspective: they often do not encompass external comparators but focus on how a system has evolved. Technical aspects influence desired outcomes in various ways (*cell 5*). For example, system usability or capacity may relate to throughput and turnaround time for testing of app notified users.

Furthermore, many outputs depend on end-user interactions. App-notified contacts are expected to follow certain procedures such as calling an infoline, getting tested, or entering quarantine (*cell 6*). Examples for such user-dependent outcome KPIs are the fraction app notified users who voluntarily enter quarantine or who seek testing. It is often instructive to express these KPIs both as yield (fraction who completes an action) and speed attributes (time until an action is completed).

Finally, the public-health-oriented metrics reflect the real-world effectiveness of DPT apps as defined above and relate to the main pandemic mitigation goals. These DPT goals are the result of the interplay between the different technical and non-technical aspects of the notification cascade. That is, the process KPIs provide information about the performance of the notification cascade required for DPT effectiveness. But KPIs do not provide direct evidence for DPT effectiveness, for which a comparator group is required (which, however, can also be a comparator without any measures such as DPT or MCT as described in section 4). In the context of the proximity contact notification step, an effectiveness evaluation could, for example, compare the time from positive testing of the index case to quarantine of the proximity contact between DPT and standard MCT.

# 6 Key considerations for the practical implementation of performance and effectiveness measures

## 6.1 Definition of expected process targets

To monitor the process performance of DPT, it is helpful to have an expectation of where indicators should stand at a given time. That is, to assess performance on the basis of absolute numbers (e.g. number of authorized uploads) or a yield (e.g. fraction of used over authorized uploads), expectations or precise benchmark targets should be defined. Because DPT is still a very novel health technology, the formulation of specific benchmarks can be challenging. What is more, targets may not only be country- but also setting- or subgroup-specific (e.g. targeting a specific app coverage in the working population).

Target definitions can in part be informed by modeling results (e.g. concerning required DPT coverage to create an effect).[1,30–32] But in many instances, only qualitative targets may be feasible because of a lack of suitable reference values. A possible approach to derive such qualitative targets is to describe "desired" effects in a hypothetical, perfectly functioning system. For example, KPIs to measure the full completion of actions, such as the fraction of authorized users that upload their information to the server (e.g., by measuring the fraction of notification codes that are redeemed). Other KPIs measuring attributes such as speed may lack a clearly defined benchmark (for example, the time from app notification to quarantine), but could be formulated in terms of a comparison to other measures (for example, the same intervals in a manual process).

## 6.2 Practical implications of the distinction between performance and effectiveness measures

The distinction between KPI and comparative effectiveness measures is more than just semantics. KPIs and effectiveness measures require different data and measurement approaches. Process metrics can be collected at different contact points in the notification cascade (Figure 1), for example at app download, during regular configuration updates, when notification codes are generated, or when notified users call an infoline. But due to the privacy-preserving, decentralized nature of DPT apps (at least those that follow the DP-3T blueprint) only provides aggregated, non-identifiable data, and these data points cannot easily be connected into a unique data stream for a specific user.

By contrast, comparative effectiveness investigations need to establish a link to processes and data collections of its comparator, which will be MCT in most settings. But from a privacy perspective, it should be clear that investigations providing unquestionable evidence for DPT effectiveness can no longer be privacy-preserving and anonymous. For example, an ideal study of DPT/MCT effectiveness should be able to connect infected cases with exposed contacts and follow their notification and quarantining cascade in great detail. One often proposed key metric is the secondary attack rate (SAR), which measures how many exposed contacts of an infected person later test PCR-positive for Sars-CoV-2.[8] But calculating precise SAR measures require an exact identification and linkage of cases and contacts, something that is not foreseen in privacy-preserving DPT apps. At the same time, DPT can also mitigate some shortcomings of MCT, for instance by improving notification speed and extending to exposed contacts who are unknown to the index case.

## 6.3 Choosing the right denominator

Selecting appropriate denominators for KPIs and public health metrics can pose challenges. While there is no universal best practice, we find that Venn diagrams can be a helpful tool to guide the search for suitable denominators. Venn diagrams are useful to illustrate the (non-)overlap between different populations of interest. In the context of DPT, these are the persons who are tested positive for Sars-Cov-2 (cases), those who are notified by DPT, and those who are informed found through manual contact tracing (MCT). DPT represents the whole population of notified app users. MCT includes all persons who were identified through manual contact tracing. The segments labeled from A-G represent population counts (e.g. corresponding to outcome-centric metrics in Figure 3).

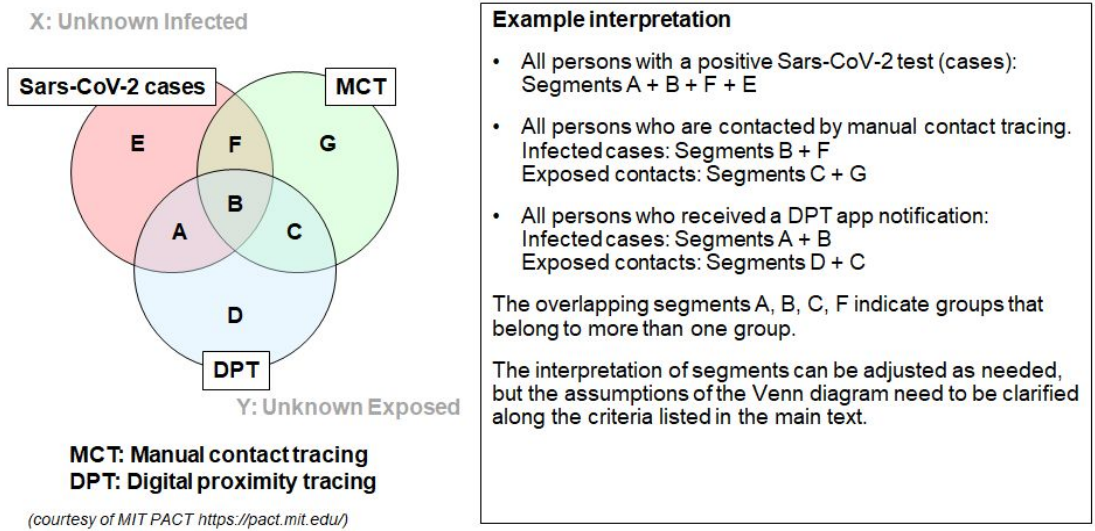

Figure 4: Example use of Venn diagrams to define denominators for KPIs and public health metrics.

While Venn diagrams may facilitate conceptual discussions, the assumptions and context must be well described. These assumptions include the following aspects.

- Goal: What type of high-level task or metric should be described by the Venn diagram?
- Population: What is the origin of the data for KPI or public health metric analysis (case series, cohort study, population-based study, administrative database)?
- Time horizon: What is the time perspective covered by the Venn diagram (cross-sectional, cumulative over a longer period)?
- Evaluation time point: At what time point are classifications into the three groups (positive tested cases, DPT, MCT) established? Shortly after the time of exposure, when PCR test results are still pending (and PCR+ denoting persons who are later confirmed to be PCR-positive? Cross-sectional at a given moment in time?
- Case definitions: Furthermore, what is the accuracy with which infection status can be determined (i.e. how to deal with infected, untested individuals)?
- Setting-specific assumptions: Finally, country-specific simplifications may be warranted based on the Test-Trace-Isolate-Quarantine strategy (e.g. whether all PCR+ cases are automatically referred to MCT). Therefore, for some countries, one can assume that segments A + E are close to 0, whereas F + B are approximating the number of positive cases.

## 6.4 Feasibility of integrating measurements directly into DPT

Given the different data requirements for KPI monitoring and effectiveness studies, the question arises how the necessary information should be collected: by integration into DPT apps and corresponding backend systems or through separate research studies?

The addition of measurement capabilities to DPT apps can be a sensitive matter. First, DPT apps following the DP-3T blueprint are not designed as data collection instruments, but as privacy-preserving notification tools that keep their users anonymous. Adding more measurement capabilities (e.g. in the backend or the app itself) leads to a data granularity-privacy trade-off. The gain in knowledge has to be weighed against a greater likelihood for de-anonymization. Adding measurement capabilities may require an increased trust by end-users in the system operators. For example, collecting exact dates of exposures, notifications, and contacts with different actors (e.g., the infoline) may, in combination, imply that study subjects may no longer be sure that their identities remain concealed. The combination of these measures may already uniquely identify persons, especially in smaller populations. If the collection of such data is to take place, it must happen transparently and app users provide informed consent. In addition, other privacy-preserving technologies can be employed that minimize the amount of data collected and limit the capability of linkage across databases[1].

The decision of whether and how to integrate additional measurements into DPT apps (beyond what is needed for notification) is one that each country needs to make separately. Such a decision must take into account specific legal considerations, overall acceptance of the DPT technology, and public expectations towards DPT privacy, as well as the individual and societal risks associated with the new data collection.

As an alternative to DPT-integrated measurements, dedicated (observational) research studies with volunteers and specifically designed databases should be considered. Given informed consent by participants, a linkage of information between DPT and MCT should be possible. For example, studies could be integrated into contact tracing and specifically survey index cases and exposed contacts regarding app usages and notifications. Other countries such as Ireland employ separate case management systems, which collect numerous complementary data that could be valuable for effectiveness research. Alternatively, app users could be presented with short questionnaires including questions regarding usage and past exposures. However, a linkage of apps with survey advertisements (even on a voluntary basis) could be regarded as intrusive and fuel privacy fears. Therefore, the advantages and disadvantages of each survey recruitment method should be deliberated carefully.

---

[1] Examples of privacy-preserving building blocks that could be used to support measurements while minimizing risks for users are multi-party computation, differential privacy, anonymous authentication, and homomorphic encryption. Other privacy technologies could also be of interest.

# Conclusion

The development of monitoring systems for DPT performance and effectiveness requires complex decisions. While there is no universal advice that could suit all settings and countries, it may help to obtain clarity on the distinctions between performance monitoring and effectiveness. Furthermore, decision makers should become aware that *not all measurements can and should be integrated into DPT apps* and connected backend systems. Separate studies or data collection systems may be needed to generate the necessary evidence for performance and effectiveness of DPT. The proposed indicator classification aims to support this process.

# Acknowledgements

The authors thank the regular participants of the DP-3T international exchange group for helpful discussions. Viktor von Wyl thanks the MIT PACT group members https://pact.mit.edu/ for sharing their insights and for providing the inspiration to use Venn diagrams.